# Hyperconjugative Effect on the Electronic Wavefunctions of Ethanol

Xiangjun Chen,[1,2,a)] Fang Wu,[1,2] Mi Yan,[1,2] Hai-Bei Li,[1,3] Shan Xi Tian,[1,3,a)]

Xu Shan,[1,2] Kedong Wang,[1,2] Zhongjun Li,[1,2] and Kezun Xu[1,2]

[1]Hefei National Laboratory for Physical Sciences at the Microscale,
[2]Department of Modern Physics,
[3]Department of Chemical Physics,
University of Science and Technology of China, Hefei, 230026, China

**Abstract**

Hyperconjugation is a basic conception of chemistry. Its straightforward effect is exhibited by the spatial delocalization characteristics of the electron density distributions or wavefunctions. Such effects on the electron wavefunctions of the highest-occupied molecular orbitals (HOMO) of two ethanol conformers are demonstrated with electron momentum spectroscopy together with natural bond orbital analyses, exhibiting the distinctly different symmetries of the HOMO wavefunctions in momentum space.

[a)] Corresponding authors. Electronic mail: xjun@ustc.edu.cn, sxtian@ustc.edu.cn

Hyperconjugation was primarily introduced by Mulliken [1, 2] nearly seventy years ago for the description of σ-extended conjugation regarding σ→π*, π→σ*, and σ→σ* donor-acceptor interactions in alkyl and other saturated substituents, particularly known from the classical conjugative interactions in aromatic molecules. The nature of hyperconjugation can be accessed by studies of its remarkable effects on molecular properties such as anomeric effect [3] and chemical reactivity and selectivity [4, 5]. Toward an understanding of the hyperconjugative effects in structural and energetic domains, the routine experiments, e.g., NMR spin-spin couplings [6, 7], vibrational shifts [8], and other experimental diagnostics [9-11], encounter some difficulties. On the other hand, the theoretically quantitative descriptions of hyperconjugative interactions can be obtained employing Weinhold's Natural Bond Orbitals (NBO) method, namely, with the so-called second-order perturbation energy $E(2)$ and spatial electron-distribution delocalization arising from charge transfer between two unperturbed NBOs [12-15]. Such processes are also closely related to the conformational preferences [13, 14]. In this Letter, a well-known fact that ethanol exists as *trans*- ($C_s$ symmetry) and *gauche*- ($C_1$ symmetry) conformers in gas-phase (as shown in the left-hand of Figure 1) [11, 16] is reexamined both with the NBO analyses and electron momentum spectroscopy (EMS) [17-19]. The hyperconjugative effects on the electron wavefunctions (i.e., electron density distributions) are clearly revealed for the two ethanol conformers.

It is difficult to rationalize the relative stability between *trans*- and *gauche*-conformers of ethanol because of the serious dependence on the theoretical methods used in calculations [9]. As shown in the lower-right of Figure 1, over the geometries optimized with the second-order many-body perturbation MP2/6-311++G(d,p) method, the results obtained with the MP2/ 6-311++G(d,p) (red solid lines) and the higher-level coupled-clusters CCSD(T)/ /6-311++G(d,p) (green solid lines) methods show that *gauche*-conformer is about 0.01 ~ 0.09 kcal/mol lower in energy than *trans*-conformer, while the hybrid density functional B3LYP//6-311++G(d,p) results (black solid lines) indicate that the latter is slightly more favorable. When the zero-point vibrational energy corrections are included, *trans*-conformer becomes more stable for these three theoretical methods (broken lines). To elucidate the hyperconjugative effects on this ambiguous stability order, the NBO analyses were performed both for these two conformers and their transition state (TS). NBO analysis transforms the canonical delocalized molecular orbitals (MOs, here are the Kohn-Sham MOs) into localized orbitals, and the hyperconjugative interaction can be treated by $E(2) = -n_\sigma F^2_{ij}/\Delta\varepsilon$, where $F_{ij}$ is the Fock matrix between the unperturbed occupied ($\sigma$) and unoccupied antibonding natural orbitals ($\sigma^*$), $n_\sigma$ is the $\sigma$ population, and $\Delta\varepsilon$ is the energy difference between the unperturbed $\sigma$ and $\sigma^*$ orbitals. For ethanol conformers, the predominant hyperconjugative interactions are shown as the schemes in the upper-right of Figure 1: two equivalent $n_O \rightarrow \sigma^*_{CH}$ interactions ($E(2) = 6.25$ kcal/mol) in *trans*-conformer, and two different interactions $n_O \rightarrow \sigma^*_{CC}$ ($E(2) = 4.93$ kcal/mol) and $n_O \rightarrow \sigma^*_{CH}$ ($E(2) = 6.73$ kcal/mol) in *gauche*-conformer, where $n_O$ represents the unperturbed occupied lone-pair orbital of oxygen. As shown in the upper-right of Figure 1, when all hyperconjugative interactions are removed, the B3LYP results show the dramatically different stability order with respect to those depicted in the low-right for *trans*-, *gauche*-, and TS conformers. However, such significant effects in energy domain are scarcely ever observed in the experiments.

Alternatively, we turn to another more straightforward effect of hyperconjugative interaction, namely, the delocalization characteristics of the electron density distribution of MO (or wavefunctions)[12, 15]. The highest-occupied MOs (HOMOs) of *trans*- and *gauche*-conformers are composed of the different spatial localized NBOs: $\varphi(trans) = 0.858n_O + 0.337\sigma_{C1H} - 0.337\sigma_{C1H}$ and $\varphi(gauche) = 0.869n_O - 0.348\sigma_{C1H} + 0.284\sigma_{C1C2}$. Such differences of the components become well-marked in momentum space. As shown in Scheme where the electron density maps in momentum space are obtained by Fourier transformation of the above spatial HOMO $\varphi$s, the distinctly different symmetries for the HOMOs of two conformers can be found for electron density maps in momentum space.

To date, one of the most powerful methods to probe the electron densities at the MO level is EMS which is an electron impact ionization experiment and whose kinematics is completely determined by detecting the two outgoing electrons in coincidence after energy and angle selection [17-19]. At the high electron impact energy, the plane wave impulse approximation provides a good description for this electron-impact single ionization process, namely, a binary (*e, 2e*) reaction. Within the target Hartree-Fock (HF) or Kohn-Sham (KS) approximation, the triple differential cross-section (TDCS) $\sigma_{EMS}$ for randomly oriented gas phase molecule can be described as $\sigma_{EMS} \propto \int d\Omega_p |\varphi_q(\mathbf{p})|^2$, where **p** is the momentum of the electron before ionization from the target and $\varphi_q(\mathbf{p})$ is the single-electron canonical HF or KS wavefunction in momentum space for the *q*th orbital from which the electron ejected. The integral is known as the spherically averaged electron momentum distribution, or electron momentum profile. Experimental momentum profile (XMP) is obtained by measuring TDCS in terms of the magnitude of the target electron momentum *p*. On the other hand, the theoretical momentum profile (TMP) is calculated by spherically averaging the square modulus of the Fourier transform of the single-electron wavefunction in position space. The nodal character of the wavefunction is preserved by a Fourier transform. So an atomic p orbital has a nodal plane through the origin in both position and momentum spaces and must yield a momentum density of zero at $p = 0$ a.u., while the momentum density for an atomic s orbital is a maximum at $p = 0$ a.u. The similar nodal characters also exist for MOs. A delocalized π MO which has a nodal plane through the bond axis yields the zero momentum density at $p = 0$ a.u., exhibiting a 'p-type' character; a fully symmetric $\sigma_g$ MO has a maximum momentum density at $p = 0$ a.u., exhibiting an 's-type' character [19].

The present EMS experiment for ethanol was carried out at an impact energy of 1200 eV + ionization potential, using a multi-channel electron momentum spectrometer [20, 21]. The instrumental energy and momentum resolution was calibrated to be about 1.2 eV and 0.1 a.u.

The ground-state electronic configurations are $(core)^6 (4a')^2 (5a')^2 (6a')^2 (7a')^2 (1a'')^2 (8a')^2 (9a')^2 (2a'')^2 (10a')^2 (3a'')^2$ for the *trans* ($C_s$ symmetry) and $(core)^6 (4a)^2 (5a)^2 (6a)^2 (7a)^2 (8a)^2 (9a)^2 (10a)^2 (11a)^2 (12a)^2 (13a)^2$ for the *gauche* ($C_1$ symmetry) conformers, respectively. Our theoretical calculations indicate that the HOMOs for both conformers have the close values of ionization potential ($3a''$: 10.99 eV, $13a$: 10.89 eV). In the experimental binding energy spectra, the lowest ionization band at 10.7 eV corresponds to the ionizations from HOMOs, i.e., $13a$ and $3a''$. Their XMP depicted in Figure 2 was obtained by plotting the band area as a function of azimuthal angles, i.e. target electron momentum p. For comparison, the theoretical momentum profiles (TMPs) were calculated with the Kohn-Sham MO wavefunctions obtained by density functional theory B3LYP and folded with the instrumental momentum resolution [22].

Since the gas-phased ethanol exists as an equilibrium mixture of one *trans* and two equivalent *gauche* conformers, the Boltzmann-weighted abundances should be taken into account in producing the TMPs. The total TMP (solid line) is the summation of the respective TMPs, 19% for *trans*- and 81% for the two equivalent *gauche*-conformers by fitting the XMP. It is not the task of the present communication to derive the accurate contributions or the relative stability of these two conformers, we pay more attention to the hyperconjugative effects on the radial (here is p) distribution of electron density (or wavefunctions) in momentum space of ethanol.

In Figure 2, one can find the different shapes of TMPs of *trans*- (dotted line) and *gauche*- (broken line) conformers. This is in line with the momentum-space density maps in Scheme, where the nodal plane of the momentum map of *trans*-conformer results in the '*p*-type' TMP with a maximum intensity at ca. p ~ 0.82 a.u. The TMP of *gauche*-conformer is the hybrid '*sp*-type' which has the highest intensity at p ~ 0 a.u. and the second maximum at p ~ 0.95 a.u. The '*sp*-type' XMP mainly arises from the same shape of momentum profile of *gauche*-conformer. The fact that the second maximum of XMP (p ~ 0.9 a.u.) slightly shifts to the lower p with respect to the TMP of *gauche*-conformer is due to the highest intensity at p~ 0.82 a.u. of the '*p*-type' TMP of *trans*-conformer. Referring the foregoing NBO decompositions of spatial MOs ($\varphi$s), we can interpret why the TMPs show the different characteristics, '*p*-type' and '*sp*-type', for *trans*- and *gauche*-conformers, respectively.

To reveal the nature of such remarkable difference of the momentum profiles between the two conformers, each components of spatial $\varphi$s are Fourier-transformed to the momentum-space and the spherically-averaged momentum density profiles are plotted in Figure 3. For 3a" of *trans*-conformer, two reverse-phased $\sigma_{C1H}$ orbitals form a pseudo–π orbital which as well as $n_O$ orbital corresponds to '*p*-type' momentum profile (see Figure 3a). In Figure 3b, the components $\sigma_{C1H}$ and $\sigma_{C1C2}$ orbitals clearly show the hybrid '*sp*-type' characteristics, yielding the total '*sp*-type' TMP of 13*a* of *gauche*-conformer. As the above-mentioned, the different delocalized components of 3a" and 13a, i.e., $\sigma_{C1H}$ and $\sigma_{C1C2}$, are totally induced by the different hyperconjugative interactions $n_O \rightarrow \sigma_{C1H}^*$ and $n_O \rightarrow \sigma_{C1C2}^*$. If these hyperconjugative interactions are shut off, the HOMOs of these two conformers should be the completely localized lone-pair $n_O$ orbitals, furthermore, the corresponding TMPs should be the typical '*p*-type'. In the present case, with aid of the NBO analyses, the '*sp*-type' XMP of ethanol definitely exhibits the hyperconjugative effects on the electron wavefunctions in momentum space.

In summary, we employed (e, 2e) EMS together with the NBO and MO analyses to demonstrate the remarkable differences of hyperconjugative interactions between two ethanol conformers. The '*sp*-type' XMP for the HOMO of ethanol stems from the delocalized electron distributions on the σ bonds induced by the hyperconjugative effect $n_O \rightarrow \sigma_{CC(or\ CH)}^*$. For the first time, such effect has been clearly observed by EMS, and the orbital 'imaging' ability of EMS is demonstrated to be a promising way to explore the complex intramolecular interactions, e.g., in this case, the hyperconjugative interactions in ethanol.

**Acknowledgements.** This research was supported by NSFC (Grant Nos. 10474090, 20673105) and MOST National Basic Research Program of China (Grant No. 2006CB922000). Professor C. E. Brion at the University of British Columbia (UBC) in Canada is acknowledged for supplying the HEMS and RESFOLD programs.

**Figure Captions:**

**Figure 1.** Left: Schematic geometries for conformers and transition state of ethanol. Right: The relative energies for conformers and transition state of ethanol (lower-right: with hyperconjugation; upper-right: without hyperconjugation). The schemes in the upper-right show hyperconjugative interactions for the two conformers of ethanol.

**Scheme.** Fourier transformation from the spatial electron density maps (left) to the density maps in momentum space (right) for the HOMOs of two conformers (upper: *trans*; lower: *gauche*).

**Figure 2.** Experimental (solid circles) and theoretical (lines) electron momentum profiles for HOMO of ethanol. Abundances of 19% for *trans* and 81% for two equivalent *gauche* conformers are considered to fit the experimental profile.

**Figure 3.** Theoretical electron momentum profiles for HOMOs and the localized $n_O$, $\sigma_{CH}$, and $\sigma_{CC}$ components of the *trans* (a) g*auche* (b) conformers.

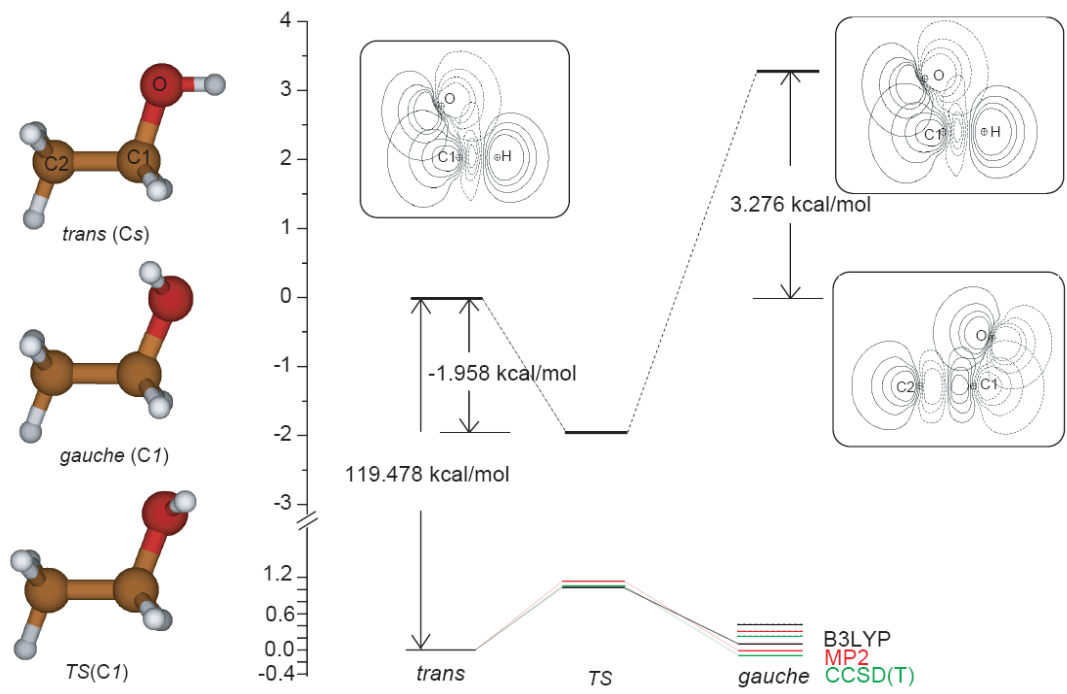

Figure1. X. J. Chen et al.

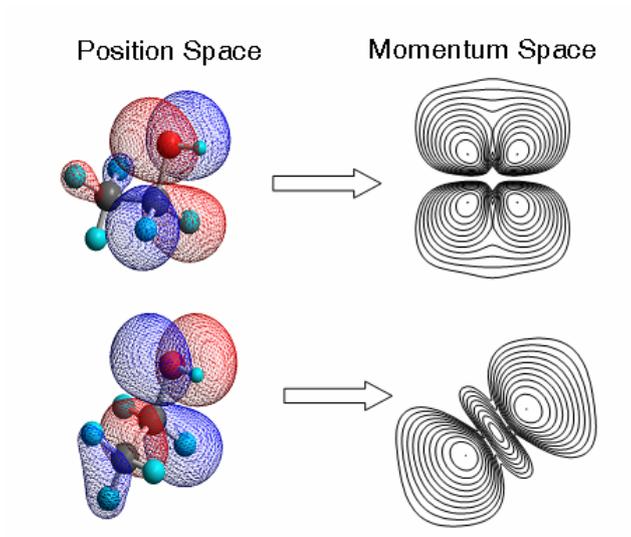

**SCHEME**  X. J. Chen et al.

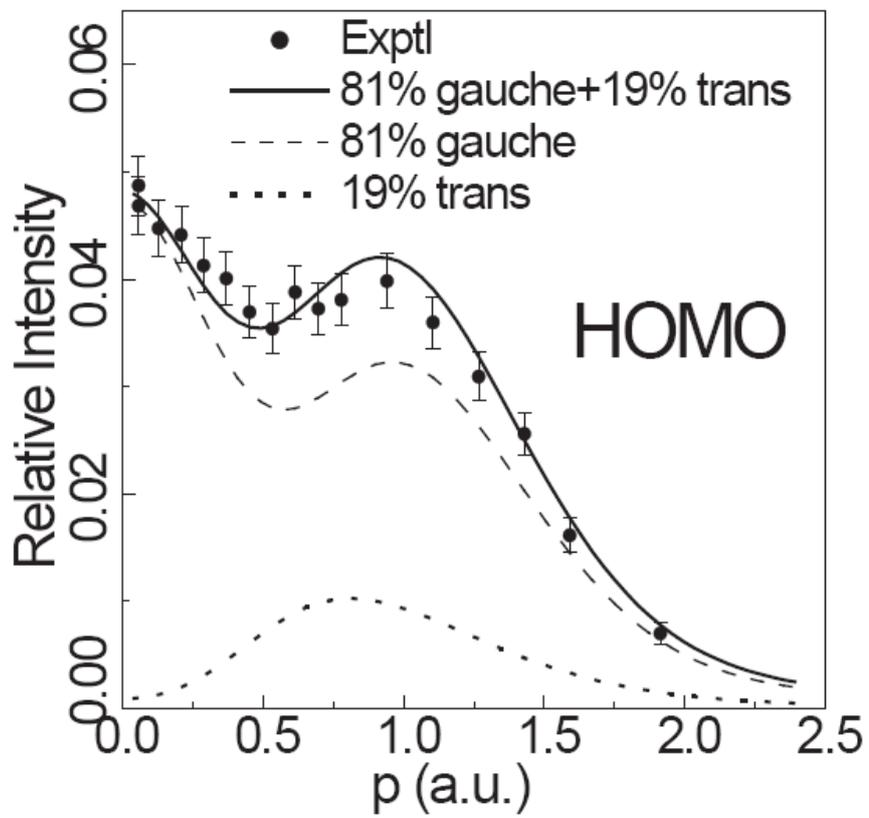

Figure2. X. J. Chen et al.

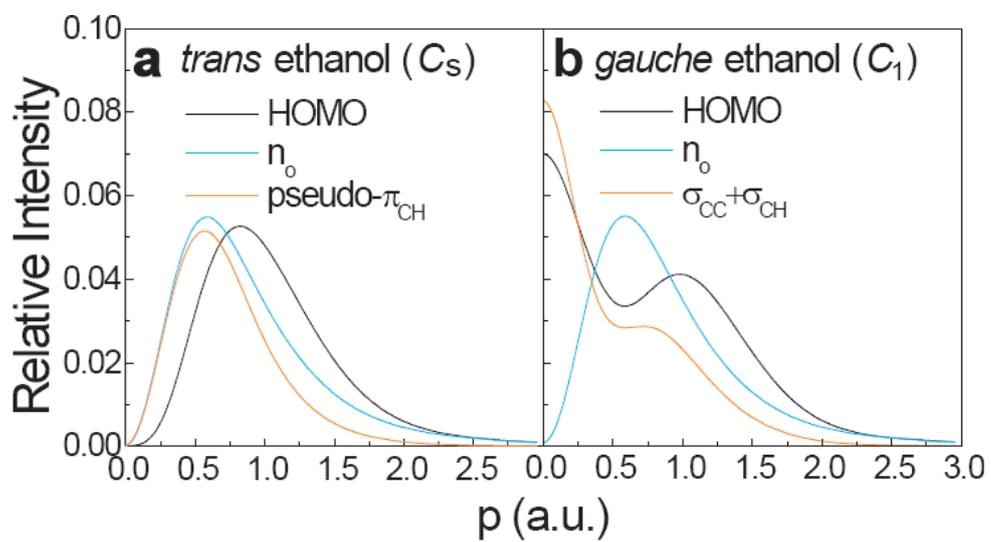

Figure3. X. J. Chen et al.